\documentclass[11pt,a4paper]{article}
\usepackage[hyperref]{acl2020}
\usepackage{times}
\usepackage{latexsym}

\usepackage{algorithm} 
\usepackage{algpseudocode} 
\usepackage{enumitem}
\usepackage{lipsum}
\usepackage{svg}
\usepackage{microtype}

\aclfinalcopy 

\title{Continuous Active Learning Using Pretrained Transformers}

\author{Nima Sadri \\
  University of Waterloo \\
  \texttt{nsadri@uwaterloo.ca} \\\And
  
  Gordon Cormack \\
  University of Waterloo \\
  \texttt{gvcormac@uwaterloo.ca} \\}

\date{}

\begin{document}
\maketitle
\begin{abstract}
Pre-trained and fine-tuned transformer models like BERT and T5 have improved the state of the art in ad-hoc retrieval and question-answering, but not as yet in high-recall information retrieval, where the objective is to retrieve substantially all relevant documents. We investigate whether the use of transformer-based models for reranking and/or featurization can improve the Baseline Model Implementation of the TREC Total Recall Track, which represents the current state of the art for high-recall information retrieval. We also introduce CALBERT, a model that can be used to continuously fine-tune a BERT-based model based on relevance feedback.
\end{abstract}

\section{Introduction}
The main objective of High recall information retrieval is to retrieve virtually all relevant documents to an information need, while minimizing the number of non-relevant documents returned. This is particularly useful is high-stake settings. For example, a lawyer might need to find all the relevant information to a specific trademark infringement, or an epidemiologist might need to find all the relevant documents to inner workings of certain protein structures. In such settings, false-positives are merely an inconvenience, but false-positives can be disastrous. 

\paragraph{} To achieve near perfect recall while minimizing the number of false-positives, there often needs to be some relevance judgment fed back to the model, which often comes in form of a human-in-the-loop. Using the example above, where the lawyer needs the documents relevant to a trademark infringement, the lawyer can start with entering a query such as "Bacardi Trademark Infringement". The system then returns one (or more) documents that it determines to be likely relevant. The user (i.e. the lawyer), would then indicate whether each returned document is relevant or not. This process would continue until the system determines that it is unlikely that there are more relevant documents that has not been shown to the user. The system can then save all the documents labeled as "relevant" and at the end of process show them to the user to read more closely. We note that, the system can also use the feedback from the user to iteratively improve the results being returned. 

\paragraph{} In fact, many law firms and corporations rely on the method mentioned above to find relevant documents from a large corpus of documents.

\section{Related-Work}
Technology Assisted Review (TAR) was first introduced by \citet{TAR} and it revolutionized how legal e-discovery is being done around the globe. The most successful implementations of TAR is a human-in-the-loop setting where the relevance feedback from user is used to improve the model. This implementation is referred to as Continuous Active Learning (CAL). The current state-of-art is an implementation by \citet{Grossman2016TREC2T}, introduced as part of the High Recall Track in TREC 2015 and 2016. This model was introduced as the baseline implementation (often referred to as BMI), but has not been beaten thus far! The model trains a logistic regression model using BM25 and TF-IDF feature vectors of the documents and the query. This simple, yet effective, model has been able to outperform complex neural network models in a variety of datasets where the goal is achieving high recall. 

\paragraph{} \citet{yang2021goldilocks} is the only attempt we are aware of where a pre-trained large scale language model, such as BERT \citep{Devlin2019BERTPO}, has been used in the CAL setting. This model initially fine-tuned BERT on the target corpus in an unsupervised fashion; then BERT was used to classify each document as relevant or not-relevant to the query, where at each iteration relevance feedback from previous iterations is used to train the model. This model, despite being much more complex than previous models, did not stand a chance when compared to BMI.

\section{Experimental Setup}

\subsection{Data}
All of our experiments are ran on a subset of the Jeb Bush Email dataset, released by \citet{Adam2015TREC} in High Recall Track of TREC in 2015. A subset of the data we use is known as \texttt{athome4} and consists of over 290,000 emails sent and received by Jeb Bush during his time as the governor of Florida. There are 34 topics of information need. We use this dataset for a few reasons. First, it is one of the few datasets where for each topic, virtually all the relevant documents are known. Second, we have access to the results for other CAL-like models on this dataset, including BMI, so we can compare our results to other models. Third, the varying lengths of the emails allows us to challenge models with limited input-lenghts.

\paragraph{} The \texttt{athome4} dataset was converted into the MSMARCO format, so it can easily be integrated into Pygaggle \footnote{pygaggle.ai/} which made it convenient to use/fine-tune the monoBERT/monoT5 models. Due to input-length limitation of BERT/T5, we had to truncate most of the documents before feeding them in BERT/T5. However, the untruncated documents were still used for BMI.

\subsection{Evaluation}
First, let us introduce the official metrics that was used to evaluate datasets in the TREC's Total Recall Track, where the \texttt{athome4} dataset was first introduced. For a topic $t$, let $R_t$ be the number of relevant documents in the corpus. The official metric used in the Total Recall Track was $Recall@4R_t+1000$ (i.e. Recall after $4R_t + 1000$ iterations). We experimented with several resource-intensive models (monoBERT, monoT5, etc.) and different hyperparameters ($K$: number of first-stage results to rerank; epochs: number of trianing epochs; etc). Due to time and resource limitations, we use $P@100$ (i.e. precision after 100 iterations) as our primary metric with the goal of finding a model that can outperform BMI on this metric. If we are able to find a model that can achieve this, we would then use $Recall@4R_t+1000$ to further explore the model).
\begin{figure*}
    \centering
    \includegraphics[width=\textwidth]{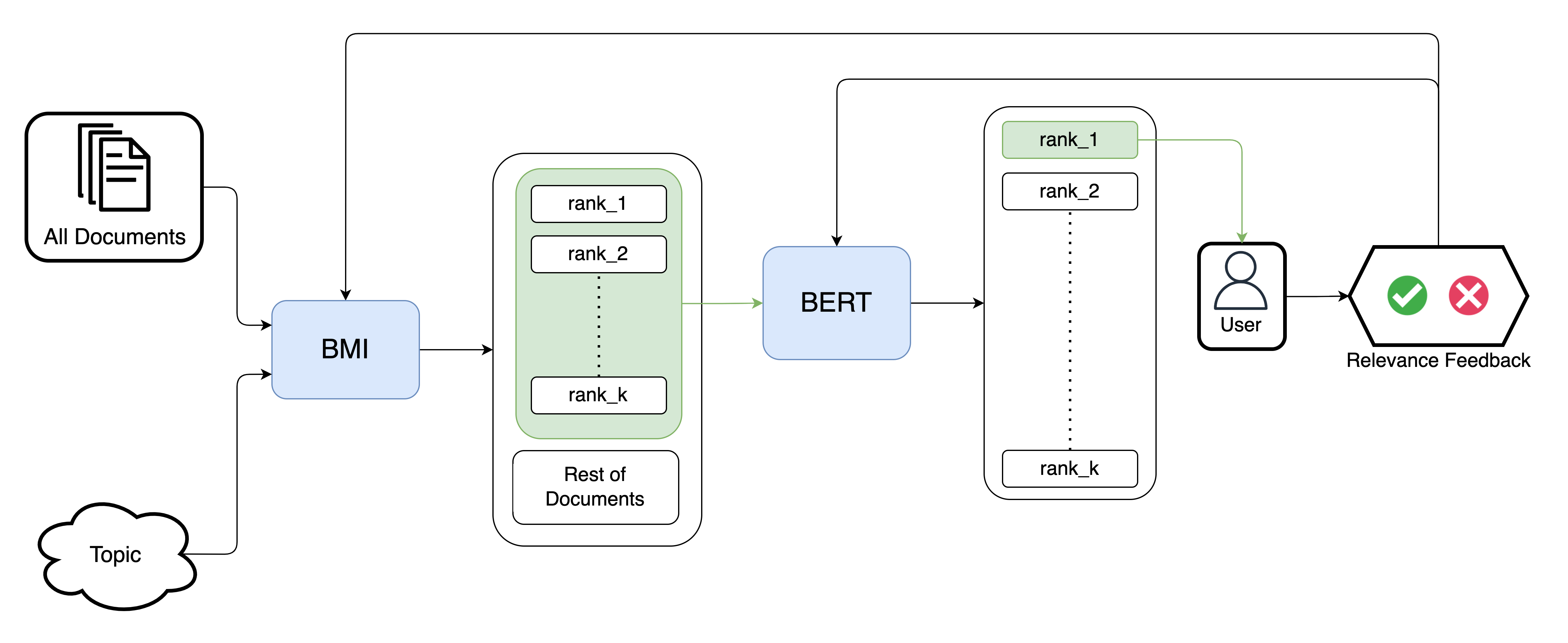}
    \caption{The CALBERT Model}
    \label{fig:calbert}
\end{figure*}

\subsection{Model}
We primarily used Transfomer-based, pre-trained language models, such as BERT \citep{Devlin2019BERTPO} and T5 \citep{raffel2019exploring}. More concretely, we started our experiments with monoBERT \citep{nogueira2019multistage} and monoT5 \citep{pradeep2021expandomonoduo} fine-tuned on the MSMARCO dataset \citep{MSMARCO}. We also tried fine-tunning monoBERT and monoT5 at each iteration based on the relevance feedback received. We note that in these cases, monoBERT and monoT5 were both utilized to rerank first-stage retrieval results recieved from BMI. We furthermore, explored creating embeddings of the documents using BERT \citep{Devlin2019BERTPO} and sentenceBERT \citep{reimers-2019-sentence-bert}. These embeddings were then used as an input to CAL in addition to the TF-IDF and BM25 feature vectors.

\subsubsection{monoBERT}
In monoBERT, given query $q$ and document $d$, the relevance score is calculated as follows \footnote{For string $s$, we use $s_n$ to denote string $s$ truncated to $n$ tokens. For example, $q_{64}$ is the query string truncated to $64$ tokens.}:
\begin{equation}
    input = "[CLS]\ q_{64} \ [SEP]\ d_{445} [SEP]"
\end{equation}
\begin{equation}
embedings = BERT(input)
\end{equation}
\begin{equation}
CLS\_emb = embedings[0]
\end{equation}
\begin{equation}
score = FullyConnected(CLS\_emb)
\end{equation}

Since BERT (and hence monoBERT) is much slower than tranditional IR models, we first rank the 290,000 documents with CAL and then rerank the top $k$ documents with monoBERT. For more details, we refer the reader to \citet{lin2021pretrained}.

\subsubsection{monoT5}
In monoT5, the query $q$ and document $d$, the relevance score is determined by computing the logits of next word being "True" or "False" when the following string is fed into T5: "Query: $q$\ \ Document: $d$\ \ Relevant: ".
Similar to monoBERT, monoT5 reranks the top $k$ results from CAL. For more details, we refer the reader to \citet{lin2021pretrained}.

\subsubsection{CALBERT}
Just as CAL trains a new logistic regression model for each topic iteratively as it receives relevance feedback, we will fine-tune a BERT model to classify documents as being relevant or non-relevant based on relevance feedback at each iteration. We name this model CALBERT. Note that, the input to CALBERT at each iteration is different documents it needs to rerank. That is, unlike monoBERT, we do not pass in the concatenation of the query and document to CALBERT. This means that we can pass in a longer portion of the document to CALBERT. More precisely, for each topic, CALBERT works as follows:
 
\begin{algorithm}
	\caption{CALBERT} 
	\begin{algorithmic}[1]
	    \State initialize CAL
	    \State initialize BERT
		\For {$iter=1,2,\ldots$}
		    \State $top_k \leftarrow CAL(docs, query, iter)$
		    \For {$doc$ in $top_k$}
		        \State $inp \leftarrow truncate("CLS" +  doc, 512)$
		        \State $emb \leftarrow BERT(inp)$
		        \State $score_{doc} \leftarrow FullyConnected(emb)$
		    \EndFor
		    \State Show argmax$_{doc}\ \{score_{doc} \}$ to user
		    \State Get relevance feedback
		    \State Fine Tune BERT
		    \State Train CAL based on feedback
		\EndFor
	\end{algorithmic} 
\end{algorithm}
Moreover, the fine-tunning process itself has some hyper-parameters. Number of epochs trained is an obvious one: after experimenting with different values, we found $5$ epochs to be yield reasonably good results. Also, at each iteration we can start with a fresh\footnote{"Fresh" here means a pretrained model that is not fine-tuned on a downstream task/corpus} BERT/T5 model and fine-tune it using all the examples in the previous iterations. This is training from "scratch". We also experimented with training models incrementally: that is, we start with a fresh model at iteration $0$ and at each iteration we train it on the relevance feedback from that iteration only. This is training "incrementally". Obviously training incrementally is much faster than training from scratch; nevertheless, we tried both to see if one has an edge over the other. We also experimented with negative sampling. Although BMI uses $100$ random samples as pseudo-negatives, our experiments showed that $100$ does not work well with Transformers. Therefore, we used a balancing technique, where at iteration $i$  we sample $\min(p_i - n_i, 0)$ pseudo-negatives where $p_i$ and $n_i$ are, respectively, the number of positive and negative relevance feedback examples we have received at iteration $i$.

\begin{table*}[t]
\centering
\begin{tabular}{lccccccc}
\hline \textbf{Model} & \textbf{k} & \textbf{epochs} & \textbf{training} & \textbf{sum} & \textbf{negatives} & \textbf{P@100} & \textbf{R@100}   \\ \hline
\textrm{BMI} & - & - & - & - & - & \textbf{81.22} & \textbf{34.98} \\
\hline
monoBERT-large & 10 & 0 & - & no & - & 72.60 & 28.60 \\
monoBERT-large & 100 & 0 & - & no & - & 61.07 & 19.15 \\
monoT5-large & 10 & 0 & - & no & - & \textbf{80.26} & \textbf{31.80} \\
monoT5-large& 100 & 0 & - & no & - & 59.73 & 21.08 \\
monoBERT-large & 10 & 0 & - & yes & - & 78.86 & 32.59 \\
monoBERT-large & 100 & 0 & - & yes & - & 64.76 & 19.77 \\
monoT5-large & 10 & 0 & - & yes & - & 78.51 & 33.02 \\
monoT5-large & 100 & 0 & - & yes & - & 65.49 & 19.99 \\
\hline
CALBERT-large & 10 & 5 & scratch & No & balanced & \textbf{80.23} & \textbf{34.68} \\
CALBERT-large & 100 & 5 & scratch & No & balanced & 76.94 & 31.92 \\
CALBERT-large & 100 & 5 & incrementally & No & balanced & 72.10 & 28.66 \\

CALRoBERTa-large & 10 & 5 & incrementally & No & balanced & 79.76 & 34.43 \\
\hline
E3-TAS & - & - & - & - & - & 76.00 & 33.59 \\
E3-all-mpnet-base-v2 & - & - & - & - & - &
\textbf{82.29} & \textbf{36.30} \\
E3-contriever & - & - & - & - & - &
58.47 & 25.79 \\

\hline
\end{tabular}
\caption{\label{results} First k documents returned by BMI, reranked using the respective models trained for the specific number of epochs. For some rows, score from BMI and the reranker is summed to get a final score – this is specified in the "sum" column.}
\end{table*}

\subsubsection{Embeddings}
This approach consisted of using embeddings generated by a Transformer-based model in place or in addition of TF-IDF and BM25 feature vectors, as an input to CAL. In particular, we explored the following inputs as inputs to CAL
\begin{enumerate}[label=E\arabic*.]
\item TF-IDF/BM25 feature vectors (Baseline)
\item Using only Transformer embeddings
\item Concatenating Transformer embeddings with TF-IDF/BM25 feature vectors
\item Training two separate CAL models, one based on TF-IDF/BM25 feature vectors, and another based on Transformer embeddings (the final score here was the sum of score from the models)
\end{enumerate}

\paragraph{} The Transformer used to generate the embedding consisted mostly of models based on BERT, fine-tuned using different data, loss functions, and techniques. The TAS (Topic Aware Sampling) model \citep{tas} model is trained on the MSMARCO dataset using a novel batch selection method. The queries are first grouped into $k$ clusters using $k$-means clustering. Then at training time, a batch of size $b$ is selected by randomly choosing $n$ topics ($n \ll k$) and $\frac{b}{n}$ queries than can be used as the basis of the training batch. We omit further details as it is beyond the scope of this work. The \verb|all-mpnet-base-v2| model is also part of the \verb|sentence-bert| library \citep{sbert}. It is based on the \verb|MPNet| model \citep{Song2020MPNetMA} and is fine-tuned with over $1$ billion sentences to generate embedding which are closer to each other in the Euclidean space; that is, embedding of similar sentences is trained to have a $\cos$ value close to $1$. The Contriever model \citep{cont} is trained using a contrastive loss to maximize "agreement" between closely located sentences, while minimizing that for sentences farther away from one another. The Contriever model uses a variety techniques for negative sampling to improve its performance; however, these details are beyond the scope of this work.

In these settings, we preprocessed all the documents and queries to generate corresponding embedding. Consequently, there was no additional latency due to BERT inference. Therefore, we did not need to conduct a rank-then-rerank approach; we simply ranked all the candidate documents at each iteration, using CAL and the appropriate embeddings (E1-E4).

\begin{table}
\centering
\begin{tabular}{ccc}
\hline \textbf{Topic} textbf{CAL} & \textbf{Transformer} \\ \hline
401 & 93.45 & 92.14 \\
402 & 98.75 & 96.71 \\
403 & 97.98 & 98.07 \\
404 & 95.05 & 95.60 \\
405 & 99.18 & 96.72 \\
406 & 94.45 & 92.13 \\
407 & 98.89 & 97.48 \\
408 & 83.62 & 80.17 \\
409 & 97.03 & 96.04 \\
410 & 1.00 & 99.55 \\
411 & 84.27 & 88.76 \\
412 & 99.22 & 98.30 \\
413 & 99.63 & 99.63 \\
414 & 1.00 & 1.00 \\
415 & 55.78 & 78.33 \\
416 & 99.59 & 99.17 \\
417 & 99.79 & 99.65 \\
418 & 95.19 & 96.79 \\
419 & 99.70 & 99.70 \\
420 & 99.59 & 99.19 \\
421 & 1.00 & 1.00 \\
422 & 1.00 & 1.00 \\
423 & 99.30 & 97.90 \\
424 & 1.00 & 1.00 \\
425 & 99.72 & 99.72 \\
426 & 98.33 & 99.17 \\
427 & 97.51 & 97.51 \\
428 & 99.14 & 98.71 \\
429 & 99.52 & 99.15 \\
430 & 97.78 & 97.98 \\
431 & 99.31 & 97.92 \\
432 & 99.29 & 97.86 \\
433 & 1.00 & 1.00 \\
434 & 1.00 & 1.00 \\
\textbf{Average} & \textbf{96.50} &	\textbf{96.77} \\
\hline
\end{tabular}
\caption{\label{pertopic} $Recall@4R_t+1000$ results for CAL and E3-all-mpnet-base-v2. This transformer was used since it performed best when compared to other transformer-based models when compared using $P@100$.}
\end{table}

\section{Results}

We report both Average Recall@100 and Average Precision@100 \footnote{"P@100" and "R@100" columns represent average precision and recall for the first 100 iterations. This notation is not to be confused by ones used in ad-hoc information retrieval evaluation, where P@100 and R@100 represent average precision and recall at the 100 cut-off. The use of this notation is justified because, unlike ad-hoc search, at each iteration we only show the user the top scoring document.} in Table \ref{results} . Note that this result is averaged over the 34 topics in the \texttt{athome4} dataset. We emphasize that all the models reranked the results from BMI in a zero-shot manner. Moreover, only the CALBERT model recieved relevance feedback; for other models, the relevance feedback was only given to the logistic regression powering BMI. For the BERT embedings methods, we only report the results for the E3 method (i.e. concatenating BERT embedding with TF-IDF/BM25 feature vectors), since this method has the most information contained in it.

\paragraph{} The baseline (BMI) performs better than most the models we explored in this work, with the exception of \verb|E3-all-mpnet-base-v2|. Even though \verb|E3-all-mpnet-base-v2| was able to beat BMI on P@100, the improvement is not statistically significant when measured using a paired T-test.

In general, fusing the first-stage ranker scores with the reranker scores seems to degrade the performance by a few percetange points, with the exception of $\textrm{monoT5}_{\textrm{large}}$ with $k = 10$. Moreoever, to our surprise, using a larger $k$ decrease precision and recall. Also, the result is inconclusive on whether monoT5 perform better than monoBERT for this task. Some embeding models, such as \verb|E3-contriever| achieved a P@100 of 0\% on some topics. We discuss the reasons behind this in the discussion section of this work.

Since \verb|E3-all-mpnet-base-v2| was our best model, we will compare it to \verb|CAL| using $Recall@4R_t+1000$. Per topic recall gain curves are included in Appendix~\ref{sec:appendix} and $Recall@4R_t+1000$ for each topic is included in Table \ref{pertopic}. Although the transformer appears to achieve superior results, a paired T-test shows that the result is not statistically significant.

\section{Discussion}
\subsection{Cold Start Problem}
Since for each topic, BMI initializes an untrained logistic regression model, it may not retrieve relevant documents accurately until it has seen a few relevant results. Parameters of BMI has been tuned so that a relevant document is retrieved in the first few iterations. However, when concatenating the Contriever embeddings with TF-IDF embeddings, the logistic regression model sometimes struggles to find any relevant documents in the first 100 iterations; this happened for $7$ of the $34$ topics. This problem can be alleviated by feeding in some relevant document to the logistic regression model before starting run for each topic; however, since in a realistic setting this is impractical, we will not utilize this trick.

\paragraph{} Another attempt to fix this cold-start problem is to use a model that does not require training (e.g. BM25) until a few relevant documents are retrieved. Since we wanted to compare our model with BMI, we decided not to utilize this technique, keeping our setting consistent with that of BMI. 

\subsection{Transformers}
We used multiple pretrained transformers to rerank the top $k$ results from BMI. However, increasing $k$ from $10$ to $100$ degraded the performance of our models significantly. This, we believe, is evidence that current state-of-the-art transformer-based models are not effective in a high-recall setting. Most advances of transformer-based models in information retrieval has been in ad-hoc search, where the goal is to simply return "some" relevant information; that is, they aim to optimize metrics such as $P@10$. This is a vastly different task than finding all relevant documents to a query. Moreover, the datasets used to fine-tune pretrained transformers for retrieval often include sparse labelings: each query has 1-2 documents labeled as relevant; MSMARCO is an example of such dataset. Therefore, models such as monoBERT and monoT5 find-tuned on MSMARCO are trained to detect a few relevant document from a candidate set of $k$ (e.g. $k = 1,000$) documents. This may be the reason why monoBERT and monoT5 cannot outperform BMI, a simple logistic regression model, when maximizing recall is the objective.

\paragraph{} Moreover, we speculate the relevance feedback received at each iteration could be used in a more effective manner to improve the results at the next iteration. Using the relevance feedback as training examples and fine-tuning the model iteratively is effective as evident by superior performance of CALBERT-large when compared to monoBERT-large. Improvements to feeding in relevance feedback to pretrained transformers can pave the path of more effective high-recall transformer-based retrievers.

\section{Conclusion}
Although pretrained transformer-based models have been making rapid progress in a wide variety of NLP tasks, their advantages in high-recall information retrieval is not yet obvious, when compared to traditional models such as logistic regression with TF-IDF feature vectors. We proposed several ways of utilizing relevance feedback in a variety of transformer-based models (such as BERT, T5, and their descendants). First, we explored with utilizing relevance feedback by fine-tunning a monoBERT/monoT5 model iteratively. Taking inspiration from CAL, we then proposed the CALBERT model, where the pretrained BERT model is fine-tunned for a specific information need; this is in contrast to the monoBERT model where the model is fine-tunned for arbitrary queries. Lastly, we generated representational embedings using a variety of transformer-based models, such as models from the sentence-bert library (e.g. mpnt), the contriver model, and the TAS model. We concatenated these emebddings with the TF-IDF feature vectors as the input to the BMI (recall that in normal BMI only the TF-IDF feature vectors are used). When $P@100$ was compared to BMI, we saw modest (though not statistically significant) improvements in one of the representational models –  the mpnt model. We ran a longer experiment with mpnt to calculate $Recall@4R_t+1000$. The gains on this longer experiment was not statistically significant either. In both the monoBERT/monoT5 and the CALBERT approaches, we used transformers to rerank the top $k$ results from BMI. As we discussed, increasing $k$ surprisingly degraded the performance, which suggests using Transformers (at least the variations used in this work) does not yield the same improvements as in the MSMARCO passage retrieval task. This, we suspect, is due to the lack of a large \textit{training} dataset in the athome4 high-recall retrieval task.

\section*{Acknowledgments}
Nima would like to thank Gordon Cormack for providing mentorship and compute resources.

\bibliography{acl2020}
\bibliographystyle{acl_natbib}

\appendix

\section{Appendices}
\label{sec:appendix}

\begin{figure}
\centering
\includegraphics[width=7.7cm]{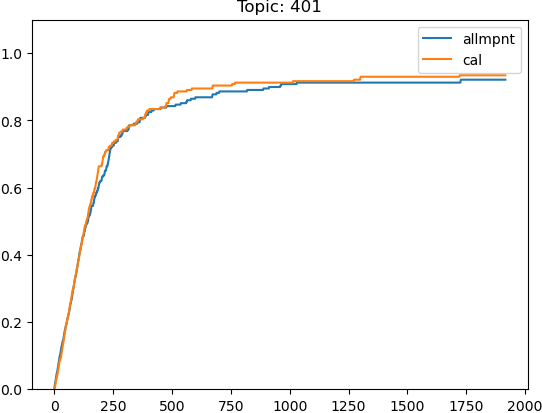}
\end{figure}
\begin{figure}
\centering
\includegraphics[width=7.7cm]{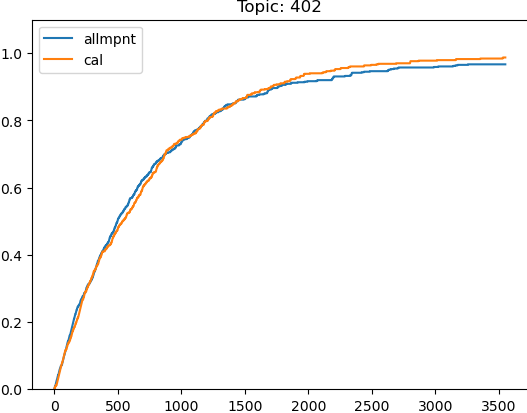}
\end{figure}
\begin{figure}
\centering
\includegraphics[width=7.7cm]{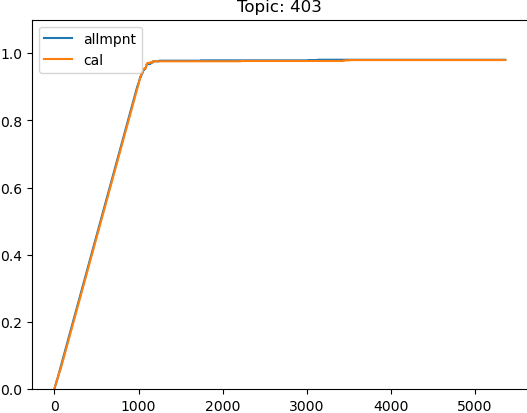}
\end{figure}
\begin{figure}
\centering
\includegraphics[width=7.7cm]{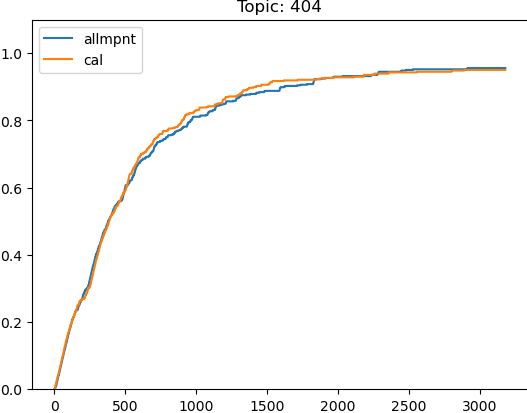}
\end{figure}
\begin{figure}
\centering
\includegraphics[width=7.7cm]{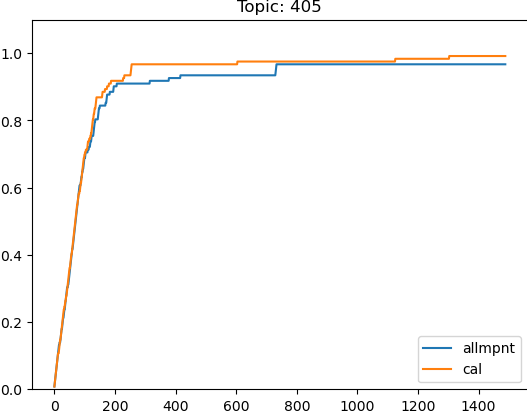}
\end{figure}
\begin{figure}
\centering
\includegraphics[width=7.7cm]{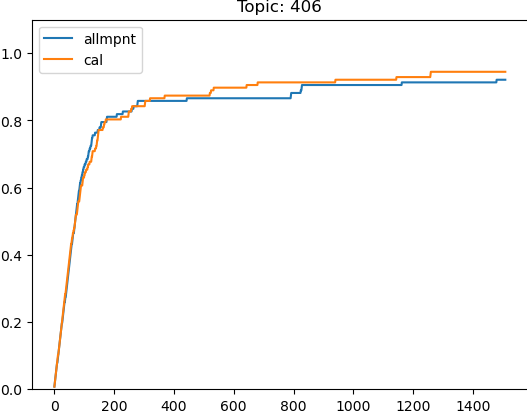}
\end{figure}
\begin{figure}
\centering
\includegraphics[width=7.7cm]{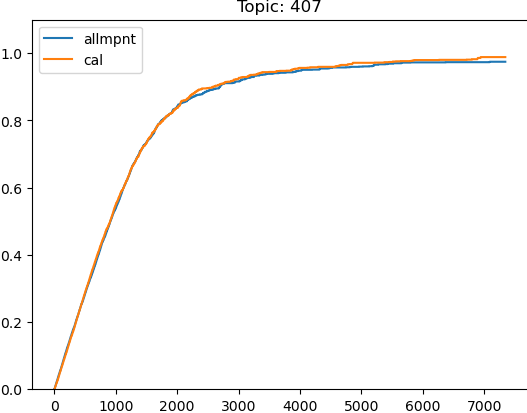}
\end{figure}
\begin{figure}
\centering
\includegraphics[width=7.7cm]{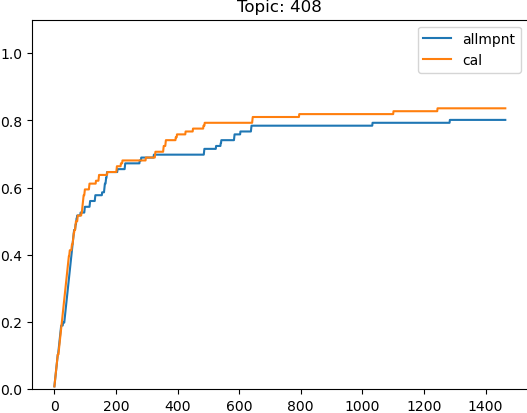}
\end{figure}
\begin{figure}
\centering
\includegraphics[width=7.7cm]{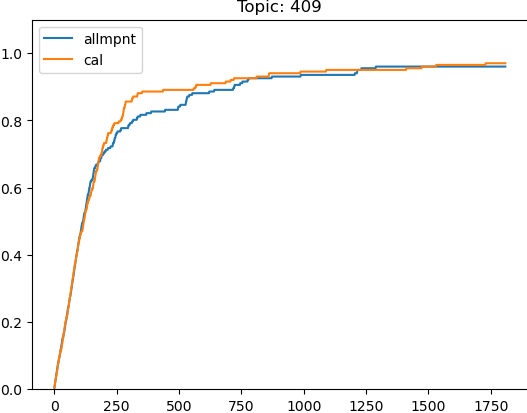}
\end{figure}
\begin{figure}
\centering
\includegraphics[width=7.7cm]{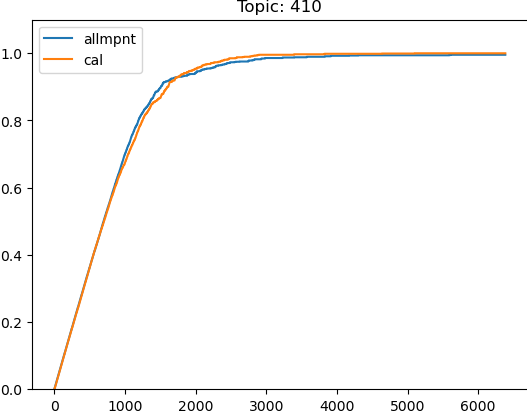}
\end{figure}
\begin{figure}
\centering
\includegraphics[width=7.7cm]{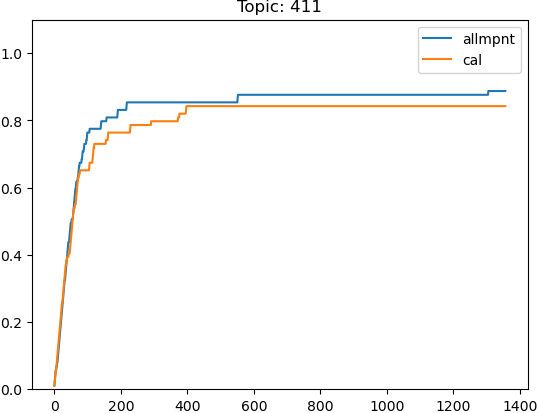}
\end{figure}
\begin{figure}
\centering
\includegraphics[width=7.7cm]{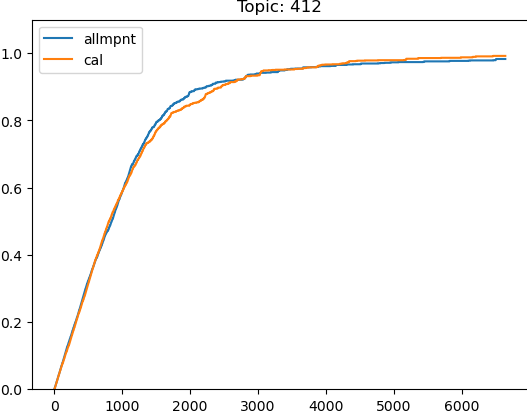}
\end{figure}
\begin{figure}
\centering
\includegraphics[width=7.7cm]{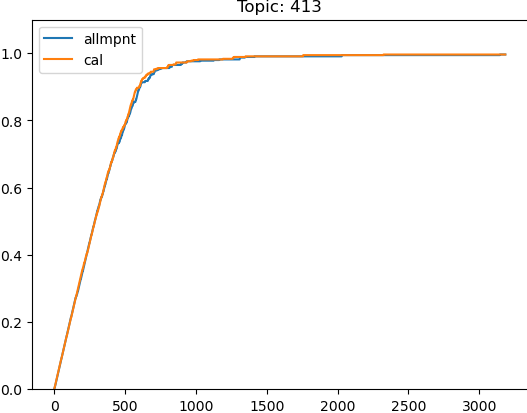}
\end{figure}
\begin{figure}
\centering
\includegraphics[width=7.7cm]{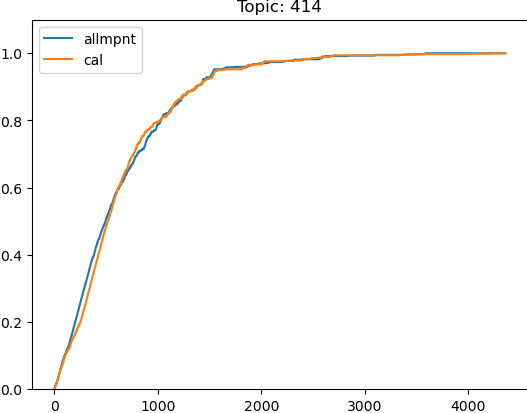}
\end{figure}
\begin{figure}
\centering
\includegraphics[width=7.7cm]{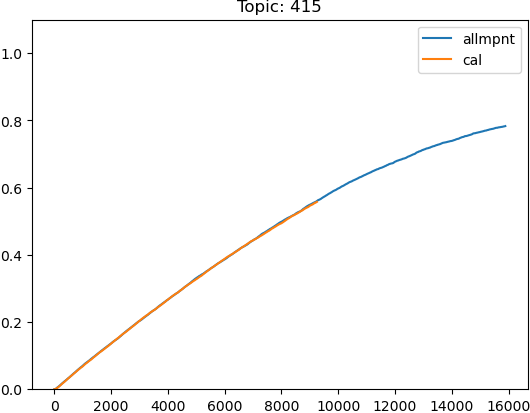}
\end{figure}
\begin{figure}
\centering
\includegraphics[width=7.7cm]{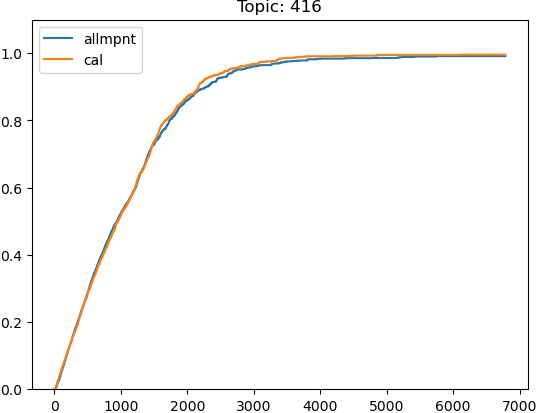}
\end{figure}
\begin{figure}
\centering
\includegraphics[width=7.7cm]{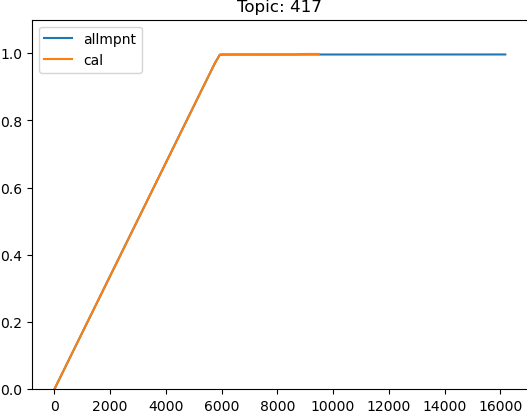}
\end{figure}
\begin{figure}
\centering
\includegraphics[width=7.7cm]{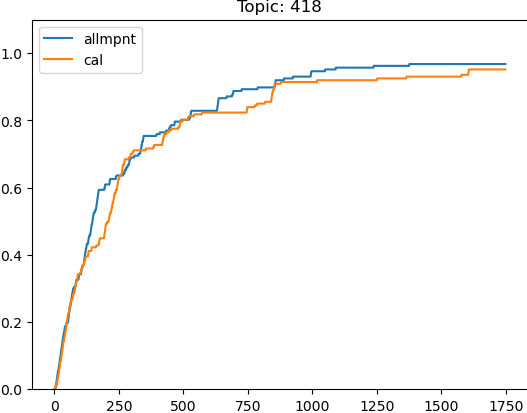}
\end{figure}
\begin{figure}
\centering
\includegraphics[width=7.7cm]{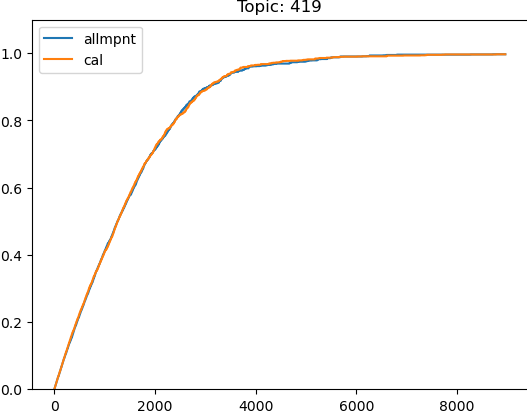}
\end{figure}
\begin{figure}
\centering
\includegraphics[width=7.7cm]{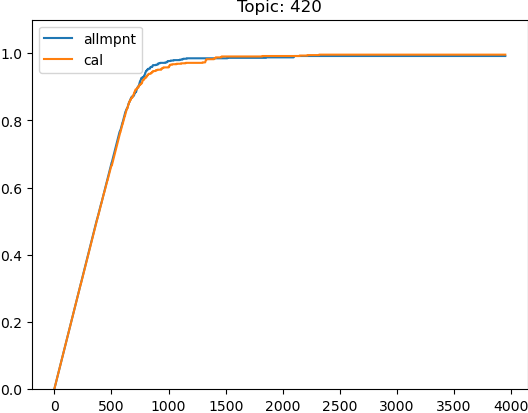}
\end{figure}
\begin{figure}
\centering
\includegraphics[width=7.7cm]{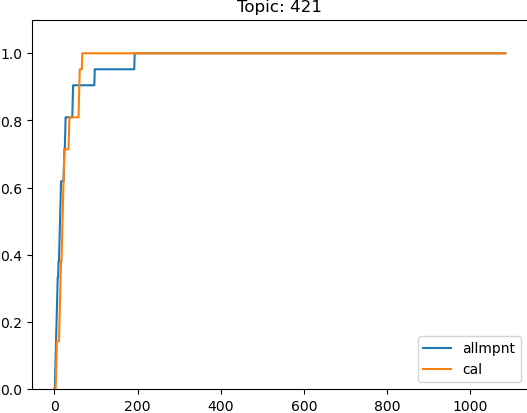}
\end{figure}
\begin{figure}
\centering
\includegraphics[width=7.7cm]{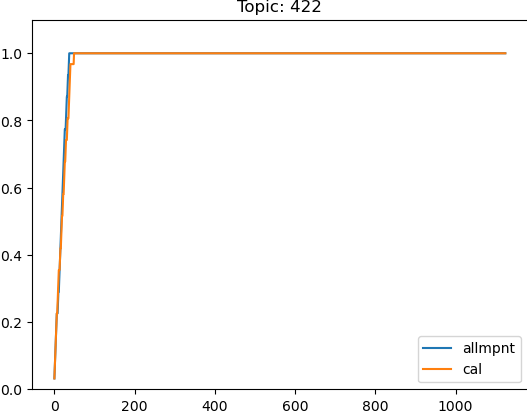}
\end{figure}
\begin{figure}
\centering
\includegraphics[width=7.7cm]{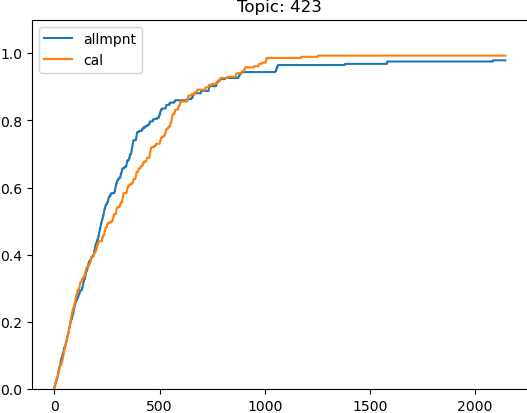}
\end{figure}
\begin{figure}
\centering
\includegraphics[width=7.7cm]{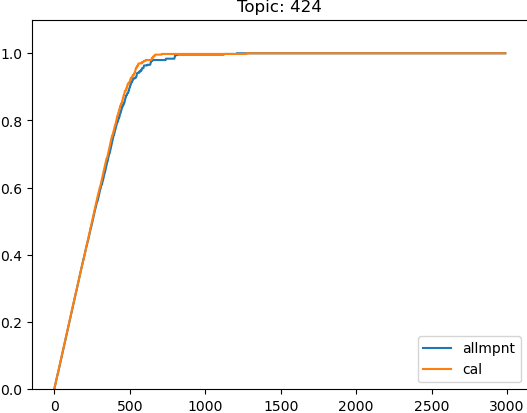}
\end{figure}
\begin{figure}
\centering
\includegraphics[width=7.7cm]{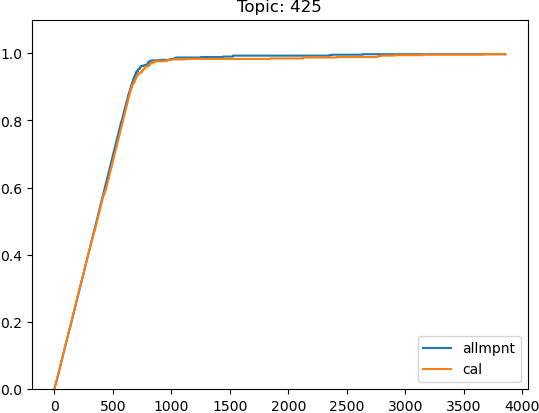}
\end{figure}
\begin{figure}
\centering
\includegraphics[width=7.7cm]{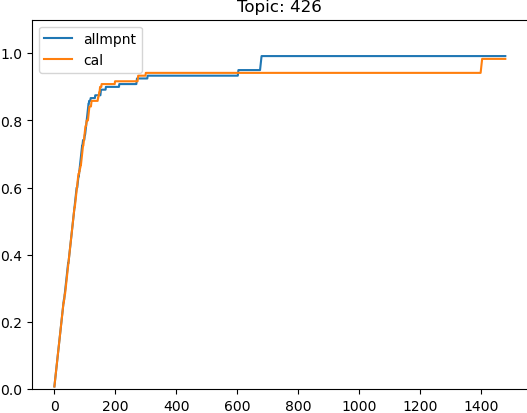}
\end{figure}
\begin{figure}
\centering
\includegraphics[width=7.7cm]{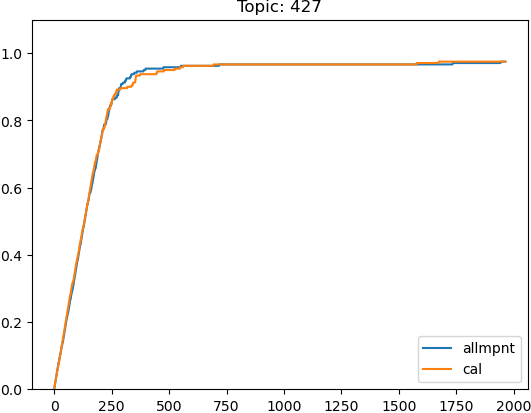}
\end{figure}
\begin{figure}
\centering
\includegraphics[width=7.7cm]{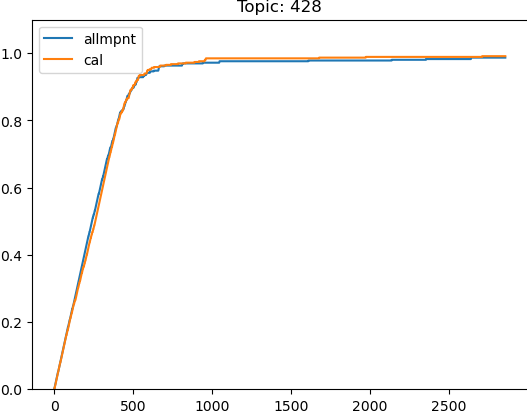}
\end{figure}
\begin{figure}
\centering
\includegraphics[width=7.7cm]{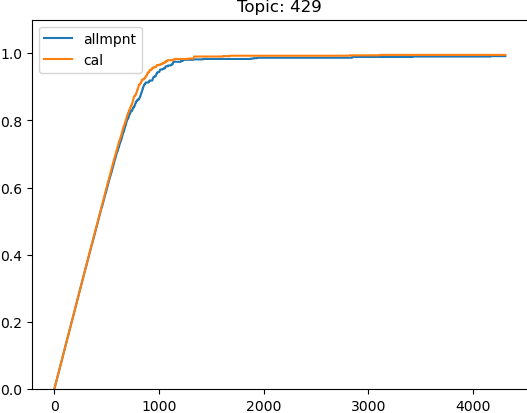}
\end{figure}
\begin{figure}
\centering
\includegraphics[width=7.7cm]{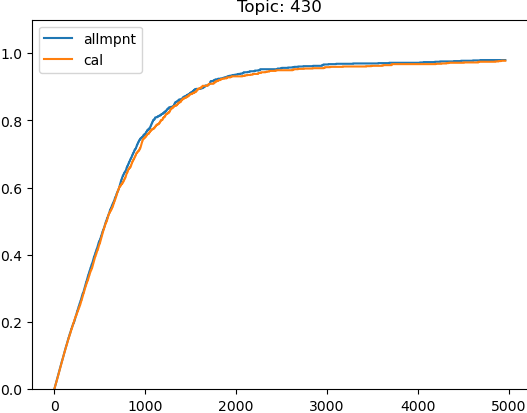}
\end{figure}
\begin{figure}
\centering
\includegraphics[width=7.7cm]{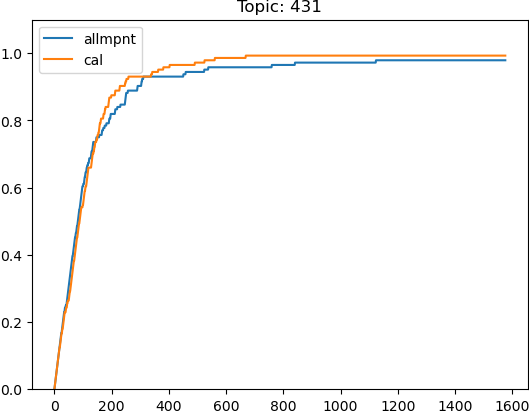}
\end{figure}
\begin{figure}
\centering
\includegraphics[width=7.7cm]{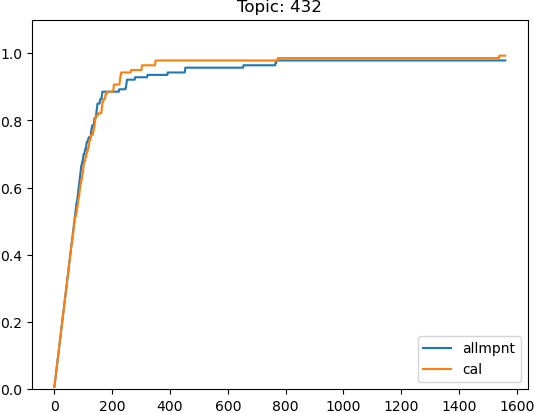}
\end{figure}
\begin{figure}
\centering
\includegraphics[width=7.7cm]{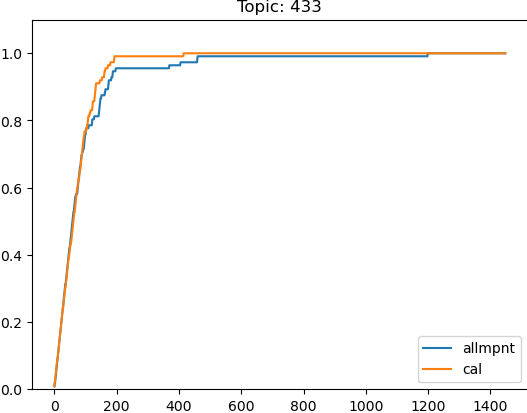}
\end{figure}
\begin{figure}
\centering
\includegraphics[width=7.7cm]{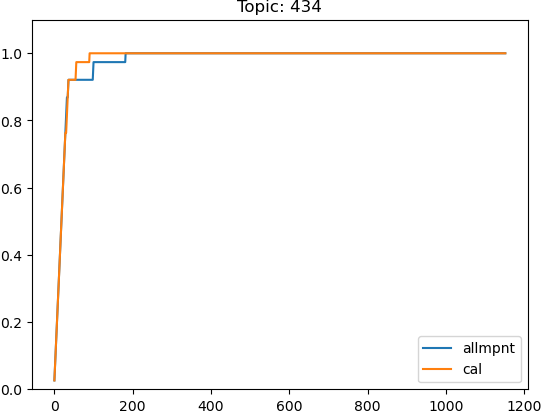}
\end{figure}

\end{document}